\documentclass[aps,pra,twocolumn,amsmath,amssymb]{revtex4}
\usepackage{times}
\usepackage{latexsym}
\usepackage{graphicx}
\usepackage{amsmath,amssymb}
\usepackage{verbatim,bbm}
\newtheorem{theorem}{Theorem}
\newtheorem{lemma}{Lemma}

\newtheorem{proposition}[theorem]{Proposition}
\newtheorem{definition}[theorem]{Definition}
\newtheorem{remark}[theorem]{Remark}

\begin{document}

\title{Continuous-variable measurement-device-independent quantum key distribution: Composable security against coherent attacks}
\author{Cosmo Lupo, Carlo Ottaviani, Panagiotis Papanastasiou, Stefano Pirandola}
\affiliation{Department of Computer Science, University of York, York YO10 5GH, UK}

\begin{abstract}
We present a rigorous security analysis of Continuous-Variable
Measurement-Device Independent Quantum Key Distribution (CV MDI QKD)
in a finite size scenario.
The security proof is obtained in two steps: by first assessing the security against
collective Gaussian attacks, and then extending to the most general class of
coherent attacks via the Gaussian de Finetti reduction.
Our result combines recent state-of-the-art security proofs for CV QKD with new findings
about min-entropy calculus and parameter estimation.
In doing so, we improve the finite-size estimate of the secret key rate.
Our conclusions confirm that CV MDI protocols allow for high rates on
the metropolitan scale, and may achieve a nonzero secret key rate against the
most general class of coherent attacks after $10^7-10^9$ quantum
signal transmissions, depending on loss and noise, and on the required level of
security.
\end{abstract}


\maketitle


\section{Introduction}\label{Sec:intro}

Quantum communication technologies, and in particular quantum key-distribution (QKD),
are rapidly progressing from research laboratories towards real-world implementations.
The ultimate goal is building a network of quantum devices
(quantum internet) enabling unconditionally secure communications
on the global scale\ \cite{QI1,QI2,QI3,QI4}.
To this end, QKD has been recently extended to a scenario where
two honest users (Alice and Bob) exploit the mediation of an
untrusted relay, operated by the eavesdropper (Eve),
to establish a secure communication channel\ \cite{MDI,CV-MDI}.
This remarkable feature is made possible by the working mechanism
of the relay itself, which activates secret correlations on the
users' remote stations by performing Bell detection on the
incoming signals and publicly announcing the results\ \cite{CV-MDI}.
This architecture has been called measurement-device independent
(MDI) QKD because, as such, the security of the communication does
not rely on the assumption that the measurement devices 
(which are more exposed to side-channel attacks than other
devices) are trusted\ \cite{MDI,CV-MDI}.

Protocols exploiting quantum continuous variables (CV)
have attracted considerable attention, for their potential of boosting
the communication rate and for their employability across mid-range
(metropolitan) distances \cite{CV-MDI,Comm}. 
The key rates achievable by CV QKD protocols
are not far from the ultimate repeater-less bound for private communication, which,
for a lossy line of transmissivity $\eta$ is
$-\log{(1-\eta)}$ bits per use\ \cite{PLOB}.
The security of CV QKD, which is very well established under Gaussian
attacks and in the asymptotic regime \cite{RMP}, 
has been recently generalized
to the most general class of coherent
attacks as well as to the finite-size setting\ \cite{Lev1,Lev2,FF1,FF2,FF3}.
In this landscape, the problem of establishing the secret key rates
achievable by CV MDI QKD in the finite-size setting has not been
yet explicitly addressed.

In this paper we fill this gap and provide a rigorous composable-security proof of the CV MDI QKD
protocol proposed in Ref.\ \cite{CV-MDI} (this proof can then be extended
to tripartite \cite{3MDI} and multipartite CV MDI protocols \cite{multik}). 
%
The security of CV MDI QKD against collective attacks can be
obtained along the lines of Ref.\ \cite{Lev1}.
Then, the extension to the most general class of coherent attacks
can be obtained by exploiting the recently introduced Gaussian de Finetti reduction\ \cite{Lev2}.
Here we apply to CV MDI QKD and improve the proof techniques of Ref.\ \cite{Lev1}:
\begin{enumerate}
\item We present a simpler analysis of parameter estimation 
that holds under general coherent attacks. Our analysis exploits the 
recently proven optimality of Gaussian attacks in the finite-size scenario \cite{Lev2}
to simplify parameter estimation.
\item We show that in CV MDI protocols, the parameter estimation
routine can be performed locally by the legitimate users with almost no
public communication.
\item We improve the secret-key rate estimates of Ref.\ \cite{Lev1} by exploiting
a new entropic inequality.
\end{enumerate}

The paper develops as follows.
We start in Section\ \ref{Sec:outline} by reviewing the CV MDI QKD protocol of Ref.\ \cite{CV-MDI}.
Section\ \ref{Sec:pe} is devoted to our new results about parameter 
estimation and its statistical analysis.
In Section\ \ref{Sec:security} we present an improved estimation
of the secret key rate obtained by applying a new entropic inequality.
A comparison with previous works is presented in Section \ref{Sec:Lev}.
To make our results more concrete, numerical examples are presented in Section\ \ref{Sec:example}.
We finally discuss the relation between security proof and experimental
realization and possible improvements in Section\ \ref{Sec:discuss}.
Finally, conclusions are presented in Section\ \ref{Sec:end}.


\section{Description of the protocol}\label{Sec:outline}

In this section, we review the CV MDI QKD protocol introduced in Ref.\ \cite{CV-MDI}.

\begin{figure}[ptb]
\centering
\includegraphics[width=0.45\textwidth]{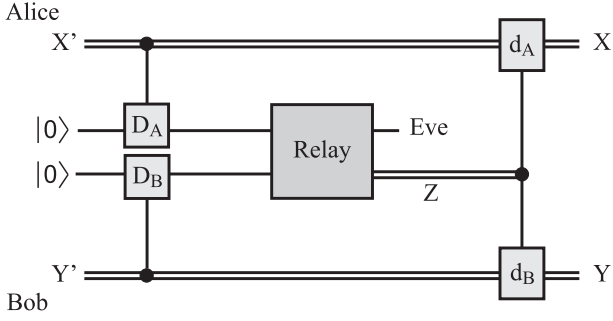}
\caption{The figure shows the scheme of the CV MDI QKD protocol
as described in details in Section \ref{Sec:outline}.
Single lines represent bosonic modes, double lines classical variables.
Time evolves from left to right.
Alice and Bob initially prepare coherent states by applying displacement
operators $D_A$, $D_B$ to the vacuum state $|0\rangle$, according to the value
of their local classical variables.
The coherent states are collected by the relay that, through some (unknown) physical
transformation, outputs a classical variable $Z$ and gives to Eve quantum
side information. Finally, Alice and Bob apply {\it classical} displacement
$d_A$, $d_B$, conditioned on the value of $Z$,
to their local classical variables.} \label{MDI-PM}
\end{figure}

The protocol develops in five steps (see Fig.\ \ref{MDI-PM}):
\begin{enumerate}

\item
{\it Coherent states preparation.}
Alice and Bob locally prepare $2 n$ coherent states,
whose complex amplitudes $\alpha' = ( q'_A + i p'_A )/2$
and $\beta' = ( q'_B + i p'_B )/2$ are drawn i.i.d.\ from
circular symmetric, zero-mean Gaussian distributions
with variance $V_M^A$ and $V_M^B$, respectively \cite{NOTAvacuum}.
The initial random variables of Alice and Bob are respectively
denoted as $X' =(q'_A,p'_A)$, $Y' =(q'_B,p'_B)$.

\item
{\it Operations of the relay.}
The $2n$ coherent states are sent to the relay. For each pair of coherent
states received the relay publicly announces a complex value
$\gamma = ( q_Z + i p_Z )/2$.

\item
{\it Parameter estimation.}
Alice and Bob estimate the covariance matrix (CM) of the variables
$( q'_A , p'_A , q'_B, p'_B, q_Z, p_Z )$.

\item
{\it Conditional displacements.}
Alice and Bob define the displaced variables
$\alpha = ( q_A + i p_A )/2$ and $\beta = ( q_B + i p_B )/2$
such that
\begin{align}
q_A & = q'_A - g_{q'_A}(\gamma)  \, , \label{dsqA} \\
p_A & = p'_A - g_{p'_A}(\gamma)  \, , \label{dspA} \\
q_B & = q'_B - g_{q'_B}(\gamma)  \, , \label{dsqB} \\
p_B & = p'_B - g_{p'_B}(\gamma)  \, , \label{dspB}
\end{align}
where $g_\star$, for each $\star = q'_A, p'_A, q'_B, p'_B$,
is an affine functions of $\gamma$.
As shown in Ref.\ \cite{io}, the optimal choice is to define the
functions as
\begin{equation}
g_\star(\gamma) = u_\star \, q_Z + v_\star \, p_Z \, ,
\end{equation}
where \cite{NOTAmean}
\begin{align}
u_\star & = \frac{ \langle \star \, q_Z \rangle \langle p_Z^2 \rangle - \langle \star \, p_Z \rangle \langle q_Z p_Z \rangle }{ \langle p_Z^2 \rangle \langle q_Z^2\rangle - \langle q_Z p_Z \rangle^2}\, , \label{u}\\
v_\star & = \frac{ \langle \star \, p_Z \rangle \langle q_Z^2\rangle - \langle \star \, q_Z \rangle \langle q_Z p_Z \rangle }{ \langle q_Z^2 \rangle \langle p_Z^2 \rangle - \langle q_Z p_Z \rangle^2  } \, . \label{v}
\end{align}
We remark that the parameters $u_\star$, $v_\star$ can be
computed directly from the estimated CM.

%
%

\item
{\it Classical post-processing.}
The variables $X= ( q_A, p_A)$, $Y=(q_B, p_B)$ represent the
local raw keys of Alice and Bob, respectively.
To conclude the protocol, the raw keys $X$, $Y$
are post-processed for error correction and privacy amplification. We assume without
loss of generality that error reconciliation is on Alice's raw key.

\end{enumerate}

The CV MDI QKD protocol described above has two main characteristic features.
The first is that Alice and Bob does not apply any measurement, as the
only measurement is performed by the untrusted relay.
This property defines the protocol as MDI\ \cite{MDI,CV-MDI}.
The second feature is that the correlations between Alice and Bob are 
generated through the variable $Z$ announced by the relay.
As explained in details in Ref.\ \cite{io}, this property
allows Alice and Bob to do parameter estimation with a negligible amount
of public communication\
\cite{NOTApergliidioti}.
Therefore, they can exploit the whole raw key for both
parameter estimation and secret key extraction.

Finally we remark that, although the variables $X$ and $Y$
have in principle infinite cardinality,
in practice they are always specified by a finite number of digits.
Furthermore, for the finite-size analysis of the protocol (as well as for
other practical issues), one needs to map the unbounded and continuous
variables $X$, $Y$ to some discrete and bounded variables $\bar X$, $\bar Y$.
The mappings $X \to \bar X$, $Y \to \bar Y$ can be realized by an Analog to Digital
Conversion (ADC) algorithm. We therefore assume that $\bar X$ and $\bar Y$ are
discrete variables with cardinality $2^{2d}$ (i.e., $d$ bits per quadrature).


\section{Parameter estimation}\label{Sec:pe}

In this Section we discuss how Alice and Bob can
estimate the CM of the variables $(q_A, p_A, q_B, p_B)$.
Without loss of generality we can assume that these variables have zero mean
and the CM has the form
\begin{align}\label{Vcano}
\mathbf{V}_{AB} = \langle \left(
\begin{array}{cccc}
q_A^2 & q_A p_A & q_A q_B & q_A p_B \\
p_A q_A & p_A^2 & p_A q_B & p_A p_B \\
q_B q_A & q_B p_A & q_B^2 & q_B p_B \\
p_B q_A & p_B q_B & p_B q_B & p_B^2 
\end{array}
\right)\rangle
= \left(
\begin{array}{cc}
x \mathbf{I} & z \mathbf{I} \\
z \mathbf{I} & y \mathbf{I}
\end{array}
\right) \, ,
\end{align}
where $\mathbf{I} = \mathrm{diag}(1,1)$, and
\begin{align}
x & = \frac{\langle q_A^2 \rangle + \langle p_A^2 \rangle }{2} \, , \label{x} \\
y & = \frac{\langle q_B^2 \rangle + \langle p_B^2 \rangle }{2} \, , \label{y} \\
z & = \frac{\langle q_A q_B \rangle + \langle p_A p_B \rangle}{2} \, . \label{z}
\end{align}


Clearly, the entries on the principal diagonal of (\ref{Vcano})
can be estimated locally by either Alice and Bob.
It remains to estimate the off diagonal term $z$.
This can be done in three different ways:
\begin{enumerate}

\item
The traditional way is that Alice and Bob exchange part of the
data via a public channel to estimate the correlation terms
$\langle q_A q_B \rangle$ and $\langle p_A p_B \rangle$.
Clearly, in order to do so they have to disclose part of the raw key,
thus reducing the final secret-key rate.
Suppose that, over a total of $n$ signals exchanged,
Alice and Bob use $m < n$ signals for parameter estimation, thus
allowing an error in the estimation of the order of $m^{-1/2}$.
Then only the remaining $n-m < n$ signals remain available for
secret key extraction (i.e., error correction and privacy amplification).

\item As noted in Ref.\ \cite{Lev1} (see also \cite{Pacher}) a rough estimate
of the signal-to-noise ratio is sufficient for Alice and Bob to run the
error correction routine before performing parameter estimation.
Then, a verification step is done to ensure that the initial estimate
was accurate enough. In this way Alice and Bob can exploit virtually
all the raw data for key generation.

\item
For our MDI protocol Alice and Bob can exploit the relations (see Section \ref{Sec:outline})
\begin{align}
q_A & = q'_A - u_{q'_A} q_Z - v_{q'_A} p_Z \, , \label{affine1} \\
p_A & = p'_A - u_{p'_A} q_Z - v_{p'_A} p_Z \, , \label{affine2} \\
q_B & = q'_B - u_{q'_B} q_Z - v_{q'_B} p_Z \, , \label{affine3} \\
p_B & = p'_B - u_{p'_B} q_Z - v_{p'_B} p_Z \, , \label{affine4}
\end{align}
to obtain
\begin{align}
z & = \frac{ \langle q_A q_B \rangle + \langle p_A p_B \rangle }{2} =\nonumber\\
& = w_1 \langle q_Z^2 \rangle + w_2 \langle p_Z^2 \rangle + w_3 \langle q_Z p_Z \rangle \, ,
\end{align}
where we have defined
\begin{align}
w_1 & := \frac{1}{2} \left( u_{q'_A} u_{q'_B} + u_{p'_A} u_{p'_B} \right) \, , \label{w1} \\
w_2 & := \frac{1}{2} \left( v_{q'_A} v_{q'_B} + v_{p'_A} v_{p'_B} \right) \, , \label{w2} \\
w_3 & := \frac{1}{2} \left( u_{q'_A} v_{q'_B} + v_{q'_A} u_{q'_B}
+ u_{p'_A} v_{p'_B} + v_{p'_A} u_{p'_B} \right) \, . \label{w3}
\end{align}
Since the variances $\langle q_Z \rangle$,
$\langle p_Z \rangle$ and the covariance $\langle q_Z p_Z\rangle$
can be locally computed by the users, then this implies that 
Alice and Bob can do parameter estimation without
publicly announcing their local data \cite{NOTApergliidioti}.
In conclusion, in this way Alice and Bob can exploit all their raw data 
for both parameter estimation and secret key extraction.

\end{enumerate}

Here we follow the latter approach because, in contrast with the 
first approach and in analogy with the second one, it requires only 
a constant (and hence negligible) amount of public communication.
Furthermore, the third approach exploits
the very structure of the MDI protocol and therefore appears
to be the most natural in this context.


\subsection{Statistical analysis of parameter estimation}

We are then left with the problem of estimating the confidence interval
associated with the statistical estimation of the CM of $(q_A,p_A,q_B,p_B)$.
It is worth stressing that this is a remarkably complex
problem in the case of general collective attacks (see Ref.\ \cite{Lev1}).
By contrast, this task becomes straightforward
under the assumption of collective Gaussian attacks.
Unlike other authors \cite{Lev2010,Jouguet2012,Usenko}, our analysis of parameter 
estimation under collective Gaussian attacks does not rely on the 
central limit theorem and is therefore mathematically rigorous in
the finite-size setting
(see instead Ref.\ \cite{Panos,China1} for a statistical analysis of
parameter estimation in CV MDI QKD that exploits the central limit theorem).

Our analysis is based on the assumption that the 
$(q'_A,p'_A,q'_B,p'_B,q_Z,p_Z)$ are Gaussian variables.
This assumption comes with no loss of generality because:
\begin{itemize}
\item The variables $(q'_A,p'_A,q'_B,p'_B)$ are Gaussian
by definition of the protocol;

\item The optimality of Gaussian attacks in the finite-size scenarion
has been established in Ref.\ \cite{Lev2}. This implies that the 
variables $(q_A,p_A,q_B,p_B)$ can be assumed to be Gaussian without loss
of generality;

\item In principle, the variables $(q_Z,p_Z)$ are not necessarily
Gaussian. Notwithstanding, by inverting Eqs.\ (\ref{affine1})-(\ref{affine4})
we can write $(q_Z,p_Z)$ as linear combinations of $(q_A,p_A,q_B,p_B)$
and $(q'_A,p'_A,q'_B,p'_B)$. Since the latter are assumed to be Gaussian,
and since a linear combination of Gaussian variables is also Gaussian,
it follows that $(q_Z,p_Z)$ are Gaussian variables too.

\end{itemize}


%

First consider the estimation of, say, $\langle q_Z^2 \rangle$, whose
estimator is the empirical variance $n^{-1} \sum_{j=1}^n q_{Zj}^2$.
Given that $q_{Zj}$ are i.i.d.\ Gaussian variables \cite{NOTAadc},
then the empirical variance is distributed (up to rescaling)
according to a chi-squared distribution.
Therefore, a confidence interval can be readily obtained
applying the cumulative distribution function of the
chi-squared distribution, or tail bounds for it.

Second, consider the estimation of the correlation
$\langle q_Z p_Z \rangle$.
We apply the identity
\begin{equation}
\langle q_Z p_Z\rangle = \frac{1}{4} \langle (q_Z + p_Z)^2 \rangle
- \frac{1}{4} \langle (q_Z - p_Z)^2 \rangle \, ,
\end{equation}
whose estimator
\begin{equation}
\frac{1}{n} \sum_{j=1}^n q_{Zj} p_{Zj} = \frac{1}{4n} \sum_{j=1}^n (q_{Zj} + p_{Zj})^2
- \frac{1}{4n} \sum_{j=1}^n (q_{Zj} - p_{Zj})^2
\end{equation}
is distributed as the sum of chi-squared variables.
Therefore, for each chi-squared variable, we can compute a confidence interval,
and then obtain a confidence interval for the quantities
$x$, $y$, and $z$ in (\ref{Vcano}) by error propagation.

An explicit calculation of the confidence intervals is
presented in Appendix \ref{App:tail}.

\section{Improved rate estimation}\label{Sec:security}

The security proof against collective or Gaussian attacks
can be obtained along the lines of Ref.\ \cite{Lev1}.
Here we present an improved estimation of the conditional
smooth min-entropy obtained by applying a new entropic
inequality.

We assume without loss of generality that the reconciliation is on Bob's
variable $\bar Y$. 
The number of (approximately) secret bits that can be extracted from the raw key is lower
bounded by the smooth min-entropy of $\bar Y$, conditioned on the
quantum state of the eavesdropper $E'$ as well as on the classical variable $Z$ \cite{Renner}:
\begin{equation}
s_{n}^{\epsilon+\epsilon_s+\epsilon_{\mathrm{EC}}}
\geq
H_{\min}^{\epsilon_s}( \bar Y | E' Z)_{ \rho^{n} } - \mathrm{leak_{EC}}(n,\epsilon_{\mathrm{EC}}) 
+ 2\log{(2\epsilon)} \, ,
\end{equation}
where we have also subtracted the information leakage
$\mathrm{leak_{EC}}(n,\epsilon_{\mathrm{EC}})$ due to error correction. The security
parameter $\epsilon+\epsilon_s+\epsilon_{\mathrm{EC}}$ comprises of three terms:
$\epsilon$ comes from the leftover hash lemma,
$\epsilon_s$ is the smoothing parameter entering the smooth conditional min-entropy, 
and $\epsilon_{\mathrm{EC}}$ is the error in the error correction routine.
%
%
Since conditioning does not increase the entropy, for any purification
$\rho^n_{ABE}$ of $\rho^n_{ABE'Z}$ we have
\begin{equation}
H_{\min}^{\epsilon_s}(\bar Y |E' Z)_{ \rho^{n} }
\geq
H_{\min}^{\epsilon_s}(\bar Y |E)_{ \rho^{n} } \, ,
\end{equation}
which implies
\begin{equation}\label{secret-bits}
s_{n}^{\epsilon+\epsilon_s+\epsilon_{\mathrm{EC}}}
\geq
H_{\min}^{\epsilon_s}(\bar Y | E)_{ \rho^{n} } - \mathrm{leak_{EC}}(n,\epsilon_{\mathrm{EC}}) 
+ 2\log{(2\epsilon)} \, .
\end{equation}


A crucial point of the security proof is the estimation of the conditional 
smooth min-entropy $H_{\min}^{\epsilon_s}( \bar Y |E )_{ \rho^{n} }$. 
Here we present an approach that yields a bound on the min-entropy that is
tighter than the one of \cite{Lev1}.
For collective (or collective Gaussian) attacks, the state $\rho^{n}$ is a tensor-power, i.e., $\rho^{n} = \rho^{\otimes n}$.
On the other hand, the state that is actually used for key generation
is the one conditioned upon error correction being successful.
Because error correction has a non-zero failure probability,
the conditional state is no longer guaranteed to be a tensor-power.
Indeed, the conditioned state has the form
\begin{equation}
\tau^{n} = p^{-1} \Pi \rho^{\otimes n} \Pi \, ,
\end{equation}
where $\Pi$ is a projector operator
(projecting on the subspace in which error correction does not abort),
and $p = \mathrm{Tr}(\Pi\rho^{\otimes n} \Pi)$ is the probability of successful
error correction.
Let us recall that the security parameter $\epsilon$ can be interpreted
as the probability that the protocol is not secure (see Appendix \ref{App:epsilon} for a review).
Therefore, the probability that the protocol is not secure, given that it
does not abort, cannot be larger than $\epsilon/p$.
This suggests a relation of the form
\begin{equation}
H_{\min}^{\epsilon}(\bar Y |E)_{ \tau^n }
\simeq H_{\min}^{p\epsilon}(\bar Y |E)_{ \rho^{\otimes n} } \, .
\end{equation}
As a matter of fact we can prove the following

\begin{theorem}\label{PS}
Given two $n-qudits$ states $\tau^n$ and $\rho^{\otimes n}$ such that
$\tau^n = p^{-1} \Pi \rho^{\otimes n} \Pi$ for some projector operator $\Pi$
and $p = \mathrm{Tr}(\Pi \rho^{\otimes n} )$, then
\begin{align}\label{new-min}
H_{\min}^{\epsilon}(\bar Y | E)_{\tau^n}
\geq H_{\min}^{\frac{2}{3}p\epsilon}(\bar X |E)_{ \rho^{\otimes n} }
+ \log{\left(  p-\frac{2}{3}p\epsilon\right)  } \, .
\end{align}
\end{theorem}
The proof is presented in Appendix \ref{App:smooth}.

Theorem \ref{PS} implies that the state can still be assumed to
be a tensor-power upon replacing
$\epsilon\rightarrow\frac{2}{3}p\epsilon$ and shortening the secret key
by $\log{\left(  p-\frac{2}{3}p\epsilon\right)  }$
bits, that is,
\begin{align}
s_{n}^{\epsilon+\epsilon_s+\epsilon_{\mathrm{EC}}}
& \geq H_{\min}^{\frac{2}{3}p\epsilon_s}(\bar Y |E)_{ \rho^{\otimes n} }
- \mathrm{leak_{EC}}(n,\epsilon_{\mathrm{EC}}) \nonumber \\
& + \log{\left(  p-\frac{2}{3}p\epsilon_s\right)  } 
+ 2\log{(2\epsilon)}
\,. \label{secret-bits-1}%
\end{align}


The conditional smooth min-entropy of the tensor-power state $\rho^{\otimes n}$
can be estimated using the Asymptotic Equipartition Property (AEP), which yields
a bound in terms of the von Neumann conditional entropy \cite{Tomamichel}:
\begin{equation}
H_{\min}^{\delta}(\bar Y | E)_{\rho^{\otimes n}}
\geq n H(\bar Y |E)_{\rho}-\sqrt{n}\,\Delta_{\mathrm{AEP}}(\delta,d)\,,\nonumber
\end{equation}
where
\begin{equation}
\Delta_{\mathrm{AEP}}(\delta,d)\leq 4(d+1)\sqrt{\log{(2/\delta^{2})}}
\end{equation}
is also a function of the dimensionality parameter $d$.

%
 
 
The next step in the security proof is to estimate the conditional entropy
\begin{equation}
H(\bar Y | E )_\rho = H(\bar Y)_\rho - I(\bar Y ; E)_\rho \, .
\end{equation}
%
%
%
%
%
Let us first consider the estimation of the mutual information $I(\bar Y ; E)_\rho$.
We remark that the latter is upper bounded by the mutual information with the
variable $Y$, i.e.,
$I(\bar Y ; E)_\rho \leq I( Y ; E )_\rho$, since the ADC algorithm cannot increase the
mutual information.
In turn, the property of extremality of Gaussian states \cite{Wolf,Raul} allows
us to write the bound $I( Y ; E )_\rho \leq I( Y ; E )_{\rho_G} \equiv I_{BE}$, 
where $\rho_G$ is a Gaussian state with same CM as $\rho$.
%
%
%
%


To conclude, we notice that the quantity
$n H(\bar{Y}) - \mathrm{leak_{EC}}(n,\epsilon_\mathrm{EC})$
is the number of (non necessarily secret) bits of common information
shared by Alice and Bob after the error correction routine.
Ideally, in the limit of large block size, ADC with arbitrarily large
precision, and perfect operations,
this quantity is expected to be equal to $n I(X;Y)_\rho$, where
$I(X;Y)$ is the mutual information between Alice and Bob.
Therefore we can put
\begin{equation}
H(\bar{Y}) - \frac{1}{n} \mathrm{leak_{EC}}(n,\epsilon_\mathrm{EC})
= \beta I(X;Y)_{\rho} \, ,
\end{equation}
where the efficiency parameter $\beta \in (0,1)$ accounts for all the
sources of non-ideality in the protocol.
The inequality $\beta I(X;Y)_\rho \geq \beta I(X;Y)_{\rho_G} \equiv \beta I_{AB}$, 
where $\rho_G$ is the Gaussian state with same first and second moments,
follows from Ref.\ \cite{Raul}. Notice that $\beta$ is also a
function of $n$ and $\epsilon_\mathrm{EC}$.

In conclusion, the results presented in this section, combined with
the security proof of \cite{Lev1}, yield the following lower bound on the 
secret key rate:
\begin{align}
& r_{n}^{ \epsilon+\epsilon_s+\epsilon_{\mathrm{EC}}+\epsilon_{\mathrm{PE}} }
= \frac{1}{n} \, s_{n}^{ \epsilon+\epsilon_s+\epsilon_{\mathrm{EC}}+\epsilon_{\mathrm{PE}} } \\
& \geq
\beta \hat I_{AB} - \hat I_{BE} 
- \frac{1}{\sqrt{n}} \, \Delta_{\mathrm{AEP}}\left(  \frac{2}{3}p\epsilon_s,d\right) \nonumber \\
& \hspace{0.5cm} + \frac{1}{n} \log{\left(  p-\frac{2}{3}p\epsilon_s\right)  }  
+ \frac{1}{n} \, 2 \log{(2\epsilon)} \, ,
\label{secret-bits-3bis}%
\end{align}
where $\hat I_{AB}$ and $\hat I_{BE}$ are the empirical estimates for the mutual
informations, and $\epsilon_\mathrm{PE}$ is the probability of error in parameter estimation.


\section{Comparison with previous security proof}\label{Sec:Lev}

Our expression for the rate in Eq.\ (\ref{secret-bits-3bis}) can be
compared to the analogous expression given in Theorem 1 of Ref.\ \cite{Lev1}.
The first difference between the two expressions is in the term
proportional to $\Delta_{\mathrm{AEP}}$ (that is the leading correction term
in our finite-size analysis), which in Ref.\ \cite{Lev1}
is replaced by \cite{NOTAtypo}
\begin{equation}
\Delta_{\mathrm{AEP}}^{(1)} = (d+1)^2 + 4 (d+1) \sqrt{\log{\frac{2}{\epsilon^2}}} + 2 \log{\frac{2}{p^2 \epsilon}}
+ 4 \frac{\epsilon d}{p\sqrt{n}} \, .
\end{equation}
It is clear that $\Delta_{\mathrm{AEP}}^{(1)} > \Delta_{\mathrm{AEP}}$,
where for small values of $p$ and $\epsilon$ the difference is dominated by
the term $2 \log{\frac{2}{p^2\epsilon}}$.
We emphasize that the fact that with our approach we obtain a smaller
finite-size correction $\Delta_{\mathrm{AEP}}$ follows from the application
of the new min-entropy inequality of Theorem \ref{PS}.

The expression for the rate in Ref.\ \cite{Lev1} also includes an additional
error term $\Delta_\mathrm{ent}$, scaling as $n^{-1/2} \log{n}$.
In our formulation this terms does no appear and has been somehow 
incorporated in the efficiency factor $\beta$.
We believe that our approach provides a better way to model
what is done in experimental implementations of the protocol.
We remark that $\Delta_\mathrm{ent}$ is the leading 
finite-size correction term in the analysis of Ref.\ \cite{Lev1}.


Finally, we exploit the Gaussian assumption to compute the confidence
intervals for parameter estimation. The result (see Appendix \ref{App:tail})
is that the elements of the CM can be estimated up to a relative error
of the order of
\begin{equation}
\sqrt{ \frac{8 \ln{(8/\epsilon_\mathrm{PR})}}{n} }
\end{equation}
with a given overall probability of error smaller than $\epsilon_\mathrm{PR}$.
This result is comparable with that of Ref.\ \cite{Lev1}: the reason
is that, although Ref.\ \cite{Lev1} considers general collective attacks,
the analysis of the parameter estimation is effectively reduced to
the Gaussian setting by applying a randomization technique.
Although we obtain finite-size correction related to parameter estimation
that are quantitatively similar to Ref.\ \cite{Lev1}, our statistical 
analysis is much simpler. This is due to the fact that we exploit the
assumption of a Gaussian attack which has been proven to come without
loss of generality even in the finite-size setting \cite{Lev2}.


%
%
%


\section{Numerical examples}\label{Sec:example}

\begin{figure}[ptb]
\centering
\includegraphics[width=0.45\textwidth]{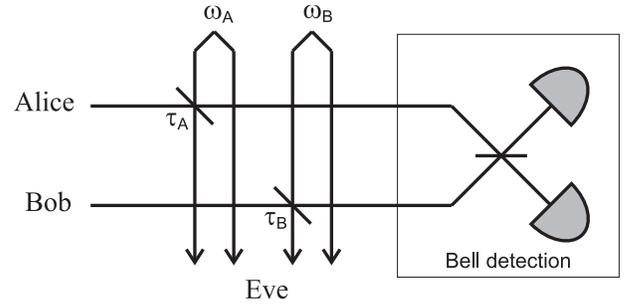}
\caption{As an example, in Section \ref{Sec:example}
we consider the case of independent entangling cloner attacks
on the two communication lines, where $\tau_A$ and $\tau_B$ are the beam-splitter
transmissivities.
The attacks also introduce independent excess noises of variances
$\xi_A = (1-\tau_A) (\omega_A-1)$, $\xi_B= (1-\tau_B)  (\omega_B-1)$.
The relay applies Bell detection on the incoming modes, whose result define the
variable $Z$ and is publicly announced.} \label{attack}
\end{figure}

The expression in Eq.\ (\ref{secret-bits-3bis}), together with the parameter estimation
analysis of Section \ref{Sec:pe},
allows us to compute the estimated secret-key directly from experimental data
for any Gaussian attack (and then extend to general attacks using the results of Ref.\ \cite{Lev2}).
In this Section, as an example, we compute the rate as function of loss and
block size for the case of an entangling cloner attack (depicted in Figure \ref{attack}).
We consider two settings:
\begin{enumerate}
\item symmetric attacks in which both communication lines from Alice to the relay
and from Bob to the relay are wiretapped with a beam-splitter with equal
transmissivity $\tau_A = \tau_B = \tau$;
\item asymmetric attacks where the relay is assumed very close to Alice station,
$\tau_A \simeq 1$.
\end{enumerate}

In both cases, following Ref.\ \cite{CV-MDI}, the eavesdropper collects all the loss from the
communication lines, and the variable $Z$ is the outcome of a perfect Bell
detection performed at the relay.
These kinds of attacks have been characterized thoroughly in Ref.\ \cite{CV-MDI}, where
the asymptotic rate (in the limit of infinite block size) has been computed as:
\begin{equation}
r^0_n = \beta \hat I_{AB} - \hat I_{BE} \, ,
\end{equation}
where the mutual informations are bounded by
the results of parameter estimation. 
%
%
In our example we choose the conservative value $\beta = 0.95$ \cite{Beta1,Beta2,Beta3,Beta4}.
(Notice that in principle the factor $\beta$ is a function of $n$ and $\epsilon_\mathrm{EC}$,
but for the sake of illustration we assume it to be constant.)

Putting
$\langle {q_A'}^2 \rangle = \langle {p_A'}^2 \rangle
= \langle {q_B'}^2 \rangle = \langle {p_B'}^2 \rangle = V_M$,
we obtain
\begin{align}
\langle q_A' q_Z \rangle & = - \sqrt{\frac{\tau_A}{2}} \, V_M \, ,\\
\langle p_A' p_Z \rangle & = \sqrt{\frac{\tau_A}{2}} \, V_M \, , \\
\langle q_B' q_Z \rangle & = \langle p_B' q_Z \rangle = \sqrt{\frac{\tau_B}{2}} \, V_M \, ,
\end{align}
and the covariances of mutually conjugate quadratures vanish.
We also have $\langle q_Z p_Z \rangle = 0$ and
\begin{equation}
\langle q_Z^2 \rangle = \langle p_Z^2 \rangle =
\frac{\tau_A + \tau_B }{2} V_M + 1 + \frac{\xi_A + \xi_B}{2}
=: \nu \, ,
\end{equation}
where $\xi_A = (1-\tau_A)(\omega_A-1)$, $\xi_B = (1-\tau_B)(\omega_B-1)$ 
are the excess noise variances and $\omega_{A,B}$ 
are the thermal noise that Eve injects in the links respectively 
(see Eq.\ (1) of Ref.\ \cite{CV-MDI}).
The only non-vanishing displacement coefficients are
\begin{align}
u_{q'_A} & = - \sqrt{ \frac{\tau_A}{2} } \, \frac{V_M}{\nu} \, , \\
v_{p'_A} & = \sqrt{ \frac{\tau_A}{2} } \, \frac{V_M}{\nu} \, , \\
u_{q'_B} & = v_{p'_B} = \sqrt{ \frac{\tau_B}{2} } \, \frac{V_M}{\nu} \, ,
\end{align}
that imply
\begin{equation}
w_1 = w_2 = - \frac{ \sqrt{\tau_A\tau_B} }{4} \, \frac{V_M^2}{\nu^2} \, ,
\end{equation}
and $w_3 = 0$.
Finally, applying Eq.\ (\ref{zmin}) we obtain
\begin{equation}
z_\mathrm{min} =
\frac{ \sqrt{\tau_A\tau_B} }{2 (1+t)} \, \frac{ V_M^2 }{\nu} \, ,
\end{equation}
and similarly, from Eq.\ (\ref{xymax}),
\begin{align}
x_\mathrm{max} & = \frac{ V_M }{1-t} \left( 1 - \frac{\tau_A}{2} \frac{V_M}{\nu} \right) \, ,\\
y_\mathrm{max} & = \frac{ V_M }{1-t} \left( 1 - \frac{\tau_B}{2} \frac{V_M}{\nu} \right) \, ,
\end{align}
with $t = \sqrt{ n^{-1} \, 8 \ln{(8/\epsilon_\mathrm{PE})} }$ (see Appendix \ref{App:tail}).

\begin{figure}[ptb]
\centering
\includegraphics[width=0.49\textwidth]{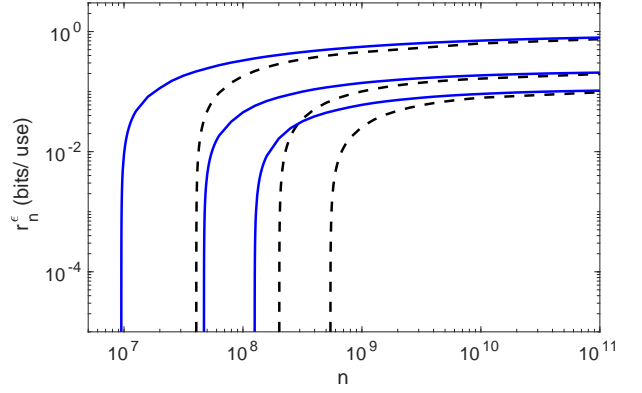}
\caption{Secret key rate vs block size for asymmetric attacks:
$\tau_A = 0.99$ and different values of $\tau_B$
(from top to bottom the attenuation of the communication line
from Bob to the relay is of $1 dB$, $2 dB$, and $4 dB$).
The excess noise is $\xi_A = 0$ and $\xi_B = 0.01$ (in shot noise unit).
Solid lines are for collective Gaussian attacks, and dashed lines
are for coherent attacks.
For both kinds of attack, the overall security parameter is smaller than $10^{-20}$.} \label{Fig:asym}
\end{figure}
\begin{figure}[ptb]
\centering
\includegraphics[width=0.49\textwidth]{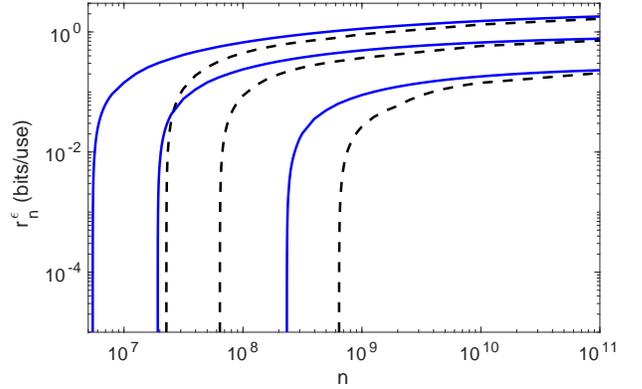}
\caption{Secret key rate vs block size for symmetric attacks
and different values of $\tau_A = \tau_B$ (from top to bottom
the symmetric attenuation is of $0.1 dB$, $0.3 dB$, $0.5 dB$, and $0.55 dB$).
The excess noise is $\xi_A = \xi_B = 0.01$ (in shot noise unit).
Solid lines are for collective Gaussian attacks, and dashed lines
are for coherent attacks.
For both kinds of attack, the overall security parameter is smaller than $10^{-20}$.} \label{Fig:sym}
\end{figure}

For collective Gaussian attacks, Eq.\ (\ref{secret-bits-3bis}) is rewritten as
\begin{align}
r_{n}^{\epsilon'}
& \geq r_n^0
- \frac{1}{\sqrt{n}} \, \Delta_{\mathrm{AEP}}\left(  \frac{2}{3}p \epsilon_s,d\right) \nonumber \\
& + \frac{1}{n} \log{\left(  p-\frac{2}{3}p  \epsilon_s\right)  }  
+ \frac{1}{n} \, 2 \log{(2\epsilon)} \, ,
\end{align}
where $\epsilon' = \epsilon+\epsilon_s+\epsilon_\mathrm{EC}+\epsilon_\mathrm{PE}$.
In Figs.\ \ref{Fig:asym}, \ref{Fig:sym} this rate is plotted vs the block size $n$,
for different values of the transmissivities and excess noise for error correction
efficiency of $\beta = 95\%$.
The plots are obtained putting $p=0.99$, 
$\epsilon = \epsilon_s = \epsilon_\mathrm{EC} = \epsilon_\mathrm{PE} = 10^{-21}$, 
hence obtaining an overall security parameter
$\epsilon' < 10^{-20}$.
We also put $d = 5$: with this choice of $d$ the error in the Shannon entropy
due to the ADC is less than $1\%$.
The rate is then obtained by maximizing over the value of modulation $V_M$.

For coherent attacks, by applying the results of Ref.\ \cite{Lev2} we obtain
\begin{align}
r_{n}^{\epsilon''} & \geq \frac{n-k}{n} \, r_n^0
- \frac{ \sqrt{n-k} }{ n } \, \Delta_{\mathrm{AEP}}\left(  \frac{2}{3}p\epsilon_s,d\right) \nonumber \\
& + \frac{1}{n} \, \log{\left(  p-\frac{2}{3}p\epsilon_s\right)  } + \frac{1}{n} \, 2 \log{(2\epsilon)}\nonumber \\
& - \frac{1}{n} \, 2 \log{ K+4 \choose 4 }
  \, ,
\end{align}
where $k$ is the number of signals used for the energy test,
$K \sim n$ and
$\epsilon'' = \frac{K^4}{50} \epsilon'$.

In Figs.\ \ref{Fig:asym}, \ref{Fig:sym} this rate is plotted vs the block size $n$,
for different values of the transmissivities and excess noise,
for error correction efficiency of $\beta = 95\%$.
The plots are obtained for
$\epsilon = \epsilon_s = \epsilon_\mathrm{EC} = \epsilon_\mathrm{PE}$
chosen in such a way to obtain $\epsilon'' < 10^{-20}$.
%
The rate is then obtained by maximizing over $k$ and the modulation $V_M$
and for $p = 0.99$.


\section{Discussion}\label{Sec:discuss}

In the case of coherent attacks, the major bottleneck limiting the rate of secret
bits generation {\it per second} comes from the classical post-processing, and in
particular the active symmetrization routine, due to the typically large size of the data set.
While it has been conjectured that such an active symmetrization might not be actually needed \cite{Lev2},
it remains an open theoretical problem to find a security proof that does not require
to perform such a computationally costly operation.

Here we present two arguments supporting the conjecture that the active symmetrization routine
may not be actually performed in any experimental realization of the protocol:

\begin{enumerate}

\item
The active symmetrization routine consists in Alice and Bob
multiplying their local raw keys by a random matrix.
Since the matrix is invertible and publicly known, such an operation cannot by any means increase the
secret key length. Therefore, we deduce that the same secret key rate might be achieved
even without performing the symmetrization routine;

\item
The symmetrization routine is also instrumental for the energy test.
After the symmetrization operation, Alice and Bob
estimate the expectation value of the energy from only a relatively small part of the raw key.
We notice that Alice and Bob can obtain an even better estimate of the mean energy
from the whole raw key. This suggests that the symmetrization step might
be avoided without affecting the energy test.

\end{enumerate}

In summary, these two arguments suggest that the requirement of performing
the symmetrization routine might be a artifact of the particular technique
used to prove the security
and therefore might not be strictly required in a practical realization of the protocol.

\section{Conclusions}\label{Sec:end}

We have presented a rigorous assessment of the
security of CV MDI QKD in the finite-size regime.
Our results are obtained by applying and modifying the results of Ref.\ \cite{Lev1},
also exploiting the Gaussian de Finetti reduction recently introduced in Ref.\ \cite{Lev2},
together with new results on parameter estimation and a new min-entropy inequality.
Because of this improvements, our estimate on the secret-key rate is
improved with respect to results of \cite{Lev1,Lev2}.

In doing this, we have shown that for our MDI protocol all the raw data can be used for both
parameter estimation and secret key extraction.
Such a unique feature is a consequence of the fact that correlations between Alice and Bob are
encoded in the variable that is publicly announced by the relay --- even though
such a variable does not contain information about the secret key (see Ref.\ \cite{io}).
It might be possible that for the same reason the security analysis of MDI QKD can be
further simplified, in particular the energy test and active symmetrization routines.
It is worth remarking that standard one-way protocols, in both direct and reverse reconciliation,
can be simulated by an MDI one, simply by assigning the relay to either Alice and Bob \cite{CV-MDI}.
For this reason, this unique property of MDI QKD can be readily extended to the one-way setting \cite{io}.

Our statistical analysis of parameter estimation is fully composable and does not rely
on the central limit theorem (and therefore is mathematically rigorous in the finite-size setting).
Notwithstanding, we do not expect that our approach gives tight bounds
on the statistical error induced by parameter estimation.
In fact, tighter bounds may be obtained following a different approach, for example
by invoking the central limit theorem as in Ref.\ \cite{Panos,China1}.

We have shown that it is in principle possible to generate secret key against
the most general class of coherent attacks for block sizes of the order of $10^7-10^9$,
depending on loss and noise, and on the required level of security.
Therefore, our results indicate that a field demonstration of CV MDI QKD
might be feasible with currently available technologies.
In particular, our composable security analysis confirms that CV MDI
protocols allow for high QKD rates on the metropolitan scale, thus
confirming the results of the asymptotic analysis first discussed in Ref.\ \cite{CV-MDI}.

{\it Note added:} after the completion of this work, other authors
have independently presented a security analysis of CV MDI QKD obtained 
by exploiting entropic uncertainty relations \cite{China2}.
Although directly applicable to obtain security against coherent attacks,
this approach is known to provide bounds on the secret key rate that in
general are not tight.

\acknowledgments
This work was supported by the
Innovation Fund Denmark within the Quantum Innovation Center Qubiz
and the
EPSRC Quantum Communications hub (EP/M013472/1).
C.L. acknowledges the valuable scientific support received from the
Quantum Physics and Information Technology Group (QPIT) of the
Technical University of Denmark (DTU), and is
specially grateful to Anthony Leverrier for insightful comments and discussions.






\appendix

\section{Operational interpretation of the security parameter}\label{App:epsilon}

Ideally, in QKD one would like to obtain a shared key that is truly random
and secret to the eavesdropper. The final state of a protocol that successfully
distributes $s$ perfectly secret bits would be represented by a density operator
of the form
\begin{equation}
\rho_0 = 2^{-s} \sum_{x=0}^{2^s-1} |x\rangle_A \langle x | \otimes |x\rangle_B \langle x | \otimes \sigma_E \, .
\end{equation}
In reality, one can only hope to get as close as possible to such an ideal scenario.
Let $\rho$ denote the final state of a given QKD protocol.
The extent to which the state $\rho$ approximates the ideal one $\rho_0$ is often
quantified in terms of the trace distance,
\begin{equation}
D(\rho,\rho_0) = \frac{1}{2} \| \rho- \rho_0\|_1 = \frac{1}{2} \mathrm{Tr}|\rho-\rho_0| \, .
\end{equation}

The trace distance has several desiderable properties for a good security
quantifier \cite{TCC,Renner,Composable}.
In particular, here we discuss its interpretation in terms of the
probability that the generated key is secret.
It is well known that the operational meaning of the trace distance is related to the
problem of quantum state discrimination \cite{Helstrom}.
Suppose one is given a black box containing either $\rho$ or $\rho_0$,
each with probability $1/2$.
Then any measurement strategy, compatible with the principles of quantum mechanics,
allows one to distinguish between the two states up to an error
probability \cite{NOTA-a}
\begin{equation}
p_e \geq \frac{1 - D(\rho,\rho_0)}{2} \, .
\end{equation}

Let us define a binary random variable $U$ with probability
distribution $P_U = (p_e,1-p_e)$.
As a matter of fact $U$ characterizes the distinguishability
of the states $\rho$ and $\rho_0$, that is, between the output of the given
QKD protocol and an ideal, perfectly secure one.
For example, if the state happens to coincide with the
ideal one, we have $P_U = P_\mathrm{sec} = ( 1/2 , 1/2 )$.
On the other hand, if the state can be perfectly distinguished
from the ideal one, $P_U = P_\mathrm{insec} = (0,1)$.

Putting $D(\rho,\rho_0) = \epsilon$ we can write
\begin{equation}
P_U = \left( \frac{1-\epsilon}{2} \, , \frac{1+\epsilon}{2} \right)
= (1-\epsilon) P_\mathrm{sec} + \epsilon P_\mathrm{insec} \, .
\end{equation}
Therefore, the probability distribution of the variable $U$ characterizing
the output of the QKD protocol is the convex sum of the probability distribution
$P_\mathrm{sec}$ associated to the ideal output state and the probability
$P_\mathrm{insec}$ associated to a state that can be perfectly distinguished from
the ideal one.
In conclusion, such a convex sum decomposition of $P_U$ allows us to interpret
$1-\epsilon$ as the probability that the output of the QKD protocol is indistinguishable from the ideal one,
and thus, for all practical purposes is itself perfectly secure.
In other words, the probability that the output of the protocol is not perfectly secure
is smaller than $\epsilon$.
Assuming the worst case scenario, below we put $\epsilon$ equal to
the probability that the key is not secret.

Taking abstraction on the state and focusing on the protocol itself,
this same reasoning is extended to the direct comparison of two protocols
$\mathcal{E}$ and $\mathcal{E}_0$, formally represented as completely positive
maps, via the diamond norm
\begin{equation}
\| \mathcal{E} - \mathcal{E}_0 \|_\diamond
= \sup_{\sigma} \| ( \mathcal{E}\otimes I - \mathcal{E}_0 \otimes I)\sigma \|_1 \, ,
\end{equation}
where the supremum is over all input states and the maps are extended
to including an ancillary system.


\section{Some properties of smooth entropy}\label{App:smooth}

One of the main tools for quantifying the security of QKD
is the conditional smooth min-entropy. In this Appendix we
review some of the main definitions and properties
(see \cite{Renner,Tomamichel} for the proofs) and derive a useful
inequality in Proposition \ref{Proposition:cutoff}
that is applied for our security proof.

\begin{definition}\label{minE}[Conditional min-entropy]
The min-entropy of $A$ conditioned on $B$ of the bipartite state $\rho_{AB}$ is
\begin{equation}
H_\mathrm{min}(A|B)_\rho := \max_\sigma \sup \left\{ \lambda : \rho_{AB} \leq 2^{-\lambda } I_A \otimes \sigma_B \right\} \, ,
\end{equation}
where $I$ is the identity operator and $\sigma$ is a subnormalized state.
\end{definition}

Here we are interested in the conditional min-entropy of
classical-quantum (CQ) states of the form
$\rho_{XB} = \sum_{x \in \mathcal{X}} P(x) | x \rangle \langle x | \otimes \omega(x)$.
In this case the conditional min-entropy can be written in terms of
the maximum guessing probability:
\begin{equation}
2^{-H_\mathrm{min}(X|B)_\rho} = \max_\mathcal{E} \sum_{x\in\mathcal{X}} P(x)
\langle x | \mathcal{E}(\omega(x)) | x \rangle \, ,
\end{equation}
where $\mathcal{E}$ is a quantum channels.

The following holds:

\begin{lemma}\label{Lemma:cutoff}
Let
$
\rho = \sum_{x \in \mathcal{X}} P(x) | x \rangle \langle x | \otimes \omega(x)
$
be a CQ state and $\mathcal{S}$ a subset of $\mathcal{X}$.
We define the projector operator $\Pi = \sum_{x \in \mathcal{S}} | x \rangle \langle x |$, and the state
$p^{-1} \Pi \rho \Pi$, with $p = \mathrm{Tr}(\Pi \rho \Pi)$.
The following inequality holds:
\begin{equation}
H_\mathrm{min} (X|B)_{ p^{-1} \Pi \rho \Pi } \geq H_\mathrm{min} (X|B)_{\rho} + \log{p} \, .
\end{equation}
\end{lemma}

{\bf Proof:} By applying the characterization of the min-entropy in terms of the guessing probability we obtain:
\begin{align}
2^{-H_\mathrm{min}(X|B)_{p^{-1} \Pi \rho \Pi}}
& = \max_\mathcal{E} \sum_{ x\in\mathcal{S} } p^{-1} P(x) \langle x | \mathcal{E}(\omega(x)) | x \rangle \\
& \leq p^{-1} \max_\mathcal{E} \sum_{x\in\mathcal{X} } P(x) \langle x | \mathcal{E}(\omega(x)) | x \rangle \\
& = p^{-1} 2^{-H_\mathrm{min}(X|B)_{\rho}} \\
& = 2^{-H_\mathrm{min}(X|B)_{\rho} - \log{p}}
\, .
\, \,  \Box
\end{align}


The smooth conditional min-entropy of $\rho$ is defined as
the maximum min-entropy in a neighborhood of $\rho$:
\begin{definition}[Smooth conditional min-entropy]
The smooth conditional min-entropy of $A$ conditioned on $B$
of the state $\rho_{AB}$ is
\begin{equation}
H^\epsilon_\mathrm{min}(A|B)_\rho := \max_{\tilde\rho} H_\mathrm{min}(A|B)_{\tilde\rho}
\end{equation}
where $\tilde\rho$ is a "smoothing state" such that
$D(\tilde\rho,\rho) \leq \epsilon$,
with $D(\tilde\rho,\rho)$ denoting the trace distance.
\end{definition}


\begin{remark}
Here we have defined the entropy smoothing
using the trace distance as in Ref.\ \cite{Renner} instead of the purified distance as done in Ref.\ \cite{Tomamichel}.
\end{remark}

\begin{remark}
For a CQ state $\rho$ it is sufficient to consider smoothing states
that are classical on the same support as $\rho$ \cite{Tomamichel}.
Therefore there exists a CQ states $\rho_\star$ such that $D(\rho_\star,\rho) \leq \epsilon$ and
\begin{equation}\label{rhostar}
H^\epsilon_\mathrm{min}(X|B)_\rho = H_\mathrm{min}(X|B)_{\rho_\star} \, .
\end{equation}
\end{remark}


\begin{lemma}\label{Lemma:Pi}
Let us consider two CQ states
$
\rho = \sum_{x \in \mathcal{X}} P(x) | x \rangle \langle x | \otimes \omega(x)
$
and
$
\rho_\star = \sum_{x \in \mathcal{X}} P_\star(x) | x \rangle \langle x | \otimes \omega_\star(x)
$
such that $D(\rho,\rho_\star) \leq \epsilon$,
and a projector operator $\Pi = \sum_{x \in \mathcal{S}} | x \rangle \langle x |$.
Then
$
D( p^{-1} \Pi \rho \Pi , p_\star^{-1} \Pi \rho_\star \Pi ) \leq \frac{3}{2} p^{-1} \epsilon \, ,
$
where
$p = \mathrm{Tr}(\Pi \rho \Pi) = \sum_{x\in\mathcal{S}} P(x)$
and
$p_\star = \mathrm{Tr}(\Pi \rho_\star \Pi) = \sum_{x\in\mathcal{S}} P_\star(x)$.
\end{lemma}
{\bf Proof.}
First notice that the trace distance between the two CQ states reads
\begin{equation}\label{D1}
D(\rho,\rho_\star) = \sum_{x\in\mathcal{X}} D( P(x)\omega(x) , P_\star(x)\omega_\star(x) ) \, ,
\end{equation}
and that $D(\rho,\rho_\star) \leq \epsilon$ implies
\begin{equation}\label{D2}
\left| p - p_\star \right| \leq \epsilon \, .
\end{equation}
We then have
\begin{align}
& D\left( p^{-1} \Pi\rho\Pi , p_\star^{-1} \Pi\rho_\star\Pi \right) \nonumber \\
& = \sum_{x\in\mathcal{S}} D\left( p^{-1} P(x) \omega(x) , p_\star^{-1} P_\star(x)\omega_\star(x) \right) \\
& \leq \sum_{x\in\mathcal{S}} D\left( p^{-1} P(x)\omega(x) , p^{-1} P_\star(x)\omega_\star(x) \right) \nonumber \\
& \hspace{1cm} + D\left( p^{-1} P_\star(x)\omega_\star(x) , p_\star^{-1} P_\star(x)\omega_\star(x) \right) \\
& = \sum_{x\in\mathcal{S}} p^{-1} D\left( P(x) \omega(x) , P_\star(x) \omega_\star(x) \right) \nonumber \\
& \hspace{1cm} + \frac{1}{2} \left| p^{-1} - p_\star^{-1} \right| P_\star(x) \| \omega_\star(x) \|_1  \\
& = \sum_{x\in\mathcal{S}}  p^{-1} D\left( P(x) \omega(x) , P_\star(x) \omega_\star(x) \right)  \nonumber \\
& \hspace{1cm} + \frac{1}{2} p^{-1} p_\star^{-1} \left| p - p_\star \right| P_\star(x)  \\
& \leq p^{-1} \epsilon + \frac{1}{2} p^{-1} \epsilon \\
& = \frac{3}{2} p^{-1}\epsilon  \, ,
\end{align}
where
in the first inequality we have applied the triangular inequality
and in the last one we have applied Eqs.\ (\ref{D1})--(\ref{D2}).
$\Box$


We are now ready to present a "smoothed" version of Lemma \ref{Lemma:cutoff}:
\begin{proposition}\label{Proposition:cutoff}
Let
$
\rho = \sum_{x \in \mathcal{X}} P(x) | x \rangle \langle x | \otimes \omega(x)
$
be a CQ state and $\mathcal{S}$ a subset of $\mathcal{X}$.
We define the projector $\Pi = \sum_{x \in \mathcal{S}} | x \rangle \langle x |$, and the (normalized) state
$\tau = p^{-1} \Pi \rho \Pi$, where $p= \mathrm{Tr}(\Pi \rho \Pi)$.
The following inequality relates the conditional smooth min-entropies
of $\rho$ and $\tau$:
\begin{equation}
H^{\epsilon}_\mathrm{min}(X|B)_{p^{-1} \Pi \rho \Pi}
\geq H^{\frac{2}{3}p\epsilon}_\mathrm{min}(X|B)_{\rho} + \log{ \left( p - \frac{2}{3} p \epsilon \right) } \, .
\end{equation}
\end{proposition}

{\bf Proof.}
Let $\rho_\star$ be a CQ state such that $D(\rho,\rho_\star) \leq \frac{2}{3} p\epsilon$.
Lemma \ref{Lemma:Pi} implies that
$D(p^{-1} \Pi \rho \Pi , p_\star^{-1} \Pi \rho_\star \Pi) \leq \epsilon$.
We then upper bound the conditional smooth min-entropy of $\tau = p^{-1} \Pi \rho \Pi$ as follows:
\begin{align}
H^{\epsilon}_\mathrm{min}(X|B)_{p^{-1} \Pi \rho \Pi}
& \geq H_\mathrm{min}(X|B)_{p_\star^{-1}\Pi \rho_\star \Pi} \\
& \geq H_\mathrm{min}(X|B)_{\rho_\star} + \log{p_\star} \\
& = H^{\epsilon'}_\mathrm{min}(X|B)_\rho + \log{p_\star} \\
& \geq H^{\epsilon'}_\mathrm{min}(X|B)_\rho + \log{( p - \epsilon' )} \, ,
\end{align}
where
in the first inequality we have applied the fact that $p_\star^{-1}\Pi \rho_\star \Pi$
is $\epsilon$-close to $p^{-1} \Pi \rho \Pi$,
in the second inequality we have applied Lemma \ref{Lemma:cutoff},
the first equality is obtained choosing a $\rho_\star$ that verifies Eq.\ (\ref{rhostar})
with $\epsilon' = \frac{2}{3} p\epsilon$,
and the last inequality is obtained from Eq.\ (\ref{D2}).
$\Box$

\subsection{Dealing with the non-zero probability that the protocol aborts}

The assumption that the state $\rho^{\otimes n}$ is a tensor
product is justified for collective attacks.
However, since error correction has non-zero probability of aborting, one should
consider the conditional probability of obtaining a secret key
given the protocol did not abort.
Unfortunately, the state conditioned on the protocol not
aborting is no longer guaranteed to have a tensor product structure.


The state $\rho^{\otimes n}$, that describes the correlations
between Bob's output measurement and Eve, is a classical-quantum (CQ) state of the form:
\begin{equation}
\rho^{\otimes n} = \sum_{x^ny^n} P(x^n, y^n) |x^n\rangle \langle x^n| \otimes |y^n\rangle \langle y^n|
\otimes \omega_E(x^n y^n) \, ,
\end{equation}
where $P(x^n, y^n)$ is the probability of a sequence of symbols $x^n, y^n$
and $\omega_E(x^n y^n)$ is the corresponding conditional state of Eve.
The protocol does not abort only on a given subset $\mathcal{S}$ of the sequences
$x^n y^n$,
therefore the state for a non-aborting protocol reads
\begin{equation}
\tau^n = p^{-1} \Pi \rho^{\otimes n} \Pi \, ,
\end{equation}
where $\Pi = \sum_{x^n y^n \in \mathcal{S}} |x^n\rangle \langle x^n| \otimes |y^n\rangle \langle y^n|$
is a projector
operator, and $p = \mathrm{Tr}(\Pi \rho^{\otimes n} \Pi)$ is the normalization factor.

Proposition \ref{Proposition:cutoff} in  Section \ref{App:smooth}
yields a simple relation between the conditional smooth
min-entropies of $\rho^{\otimes n}$ and $\tau^n$, namely
\begin{equation}
H_\mathrm{min}^\epsilon (X|E)_{\tau^n} \geq H_\mathrm{min}^{\frac{2}{3}p\epsilon} (X|E)_{\rho^{\otimes n}} + \log{ \left( p-\frac{2}{3}p\epsilon \right) } \, ,
\end{equation}
where $p$ is interpreted as the probability that the protocol does no abort.





\section{Tail bounds}\label{App:tail}

The cumulative distribution function of the chi-squared variable $\chi^2(k)$
with $k$ degrees of freedom is
$F(x;k) = \frac{ \Gamma[k/2,x/2] }{ \Gamma[k/2]}$, where
$\Gamma[k/2]$ is the Euler Gamma function, and $\Gamma[k/2,x/2]$ is the lower incomplete Gamma function.

To bound the cumulative distribution function we can use, for example, the tail bounds:
\begin{align}
\mathrm{Pr} \left\{ k < \frac{\chi}{1+t} \right\} & < e^{- n t^2/8} \, , \\
\mathrm{Pr} \left\{ k > \frac{\chi}{1-t} \right\} & < e^{- n t^2/8} \, .
\end{align}
(These bounds are derived from the Chernoff bound using the fact that
distribution of $\chi^2(k)$ is sub-exponential with parameters $(2 \sqrt{k} , 4)$).

A direct application of these bounds yields
\begin{align}
\mathrm{Pr} \left\{  \langle q^2_Z \rangle < \frac{n^{-1} \sum_j q_{Zj}^2}{1+t} \right\}
& \leq e^{- n t^2/8} \, , \\
\mathrm{Pr} \left\{  \langle q^2_Z \rangle > \frac{n^{-1} \sum_j q_{Zj}^2}{1-t} \right\}
& \leq e^{- n t^2/8} \, ,
\end{align}
together with similar bounds for the quantities $\langle p_Z^2 \rangle$,
$\langle q_A^2 \rangle$, $\langle p_A^2 \rangle$
$\langle q_B^2 \rangle$, $\langle p_B^2 \rangle$.

\begin{widetext}

We also obtain
\begin{align}
& \mathrm{Pr} \left\{ \langle q_Z p_Z \rangle
> \frac{n^{-1} \sum_j (q_{Zj}+p_{Zj})^2}{4(1-t)} - \frac{n^{-1} \sum_j (q_{Zj}-p_{Zj})^2}{4(1+t)}
\right\} \nonumber \\
& \leq \mathrm{Pr} \left\{ \langle ( q_Z + p_Z)^2 \rangle
> \frac{n^{-1} \sum_j (q_{Zj}+p_{Zj})^2}{(1-t)} \right\}
+ \mathrm{Pr} \left\{ \langle ( q_Z - p_Z)^2 \rangle
< \frac{n^{-1} \sum_j (q_{Zj}-p_{Zj})^2}{(1+t)} \right\} \nonumber \\
& \leq 2 e^{- n t^2/8} \, ,
\end{align}
and analogously
\begin{align}
\mathrm{Pr} \left\{ \langle q_Z p_Z \rangle
< \frac{n^{-1} \sum_j (q_{Zj}+p_{Zj})^2}{4(1+t)} - \frac{n^{-1} \sum_j (q_{Zj}-p_{Zj})^2}{4(1-t)}
\right\}
\leq 2 e^{- n t^2/8} \, .
\end{align}

This implies
\begin{align}
\mathrm{Pr}\left\{ x > x_\mathrm{max} \right\} \leq 2 e^{- n t^2/8} \, , \, \,
\mathrm{Pr}\left\{ y > y_\mathrm{max} \right\} \leq 2 e^{- n t^2/8} \, , \, \,
\mathrm{Pr}\left\{ z < z_\mathrm{min} \right\} \leq 4 e^{- n t^2/8} \, ,
\end{align}
with
\begin{align}\label{xymax}
x_\mathrm{max} =  \frac{1}{1-t} \sum_j \frac{ q_{Aj}^2 + p_{Aj}^2}{2n} \, , \, \,
y_\mathrm{max} =  \frac{1}{1-t} \sum_j \frac{ q_{Bj}^2 + p_{Bj}^2}{2n} \, ,
\end{align}
and
\begin{equation}\label{zmin}
z_\mathrm{min} = \hspace{-0.2cm}  \min_{s_1,s_2,s_3 \in \{-1,1\}}
\left| w_1 \frac{n^{-1} \sum_j q_{Zj}^2}{1+ s_1 t}
+ w_2 \frac{n^{-1} \sum_j p_{Zj}^2}{1+s_2 t}
 + w_3 \left( \frac{n^{-1} \sum_j (q_{Zj}+p_{Zj})^2}{4(1+ s_3 t)} - \frac{n^{-1} \sum_j (q_{Zj}-p_{Zj})^2}{4(1-s_3 t)} \right)
\right| \, ,
\end{equation}
where $w_1$, $w_2$ and $w_3$ are defined in Eqs.\ (\ref{w1})-(\ref{w3}).

For example, putting
\begin{equation}
t = \sqrt{ \frac{8 \ln{(8/\epsilon_\mathrm{PE})}}{n} }
\end{equation}
we finally obtain
\begin{equation}
\mathrm{Pr} \left\{ x > x_\mathrm{max} \vee y > y_\mathrm{max} \vee z < z_\mathrm{min} \right\}
\leq \epsilon_\mathrm{PE} \, .
\end{equation}

\end{widetext}


\begin{thebibliography}{9}

\bibitem{QI1}
H. J. Kimble, Nature \textbf{453}, 1023 (2008).

\bibitem{QI2}
S. Pirandola and S. L. Braunstein,
Nature \textbf{532}, 169 (2016).

\bibitem{QI3}
S. Pirandola \textit{et al., }
Nature Photon. \textbf{9}, 641 (2015).

\bibitem{QI4}
U. L. Andersen, J. S. Neergaard-Nielsen, P. van Loock, A. Furusawa,
Nature Phys. \textbf{11}, 713 (2015).

\bibitem {MDI}
S. L. Braunstein, S. Pirandola, Phys. Rev. Lett. \textbf{108},
130502 (2012); H.-K. Lo, M. Curty, B. Qi, Phys. Rev. Lett. \textbf{108},
130503 (2012).

\bibitem {CV-MDI}
S. Pirandola, C. Ottaviani, G. Spedalieri, C. Weedbrook, S.
L. Braunstein, S. Lloyd, T. Gehring, C. S. Jacobsen, U. L. Andersen,
Nature Photon. {\bf 9}, 397 (2015).

\bibitem{Comm}
S. Pirandola, C. Ottaviani, C. S. Jacobsen, G. Spedalieri, S.
L. Braunstein, S. Lloyd, T. Gehring, U. L. Andersen,
Nature Photon. \textbf{9}, 776 (2015).

\bibitem{PLOB}
S. Pirandola, R. Laurenza, C. Ottaviani, L. Banchi,
Nature Commun. {\bf 8}, 15043 (2017).
See also arXiv:1510.08863 (2015).

\bibitem{RMP}
C. Weedbrook, S.Pirandola, R. Garc\'{\i}a-Patr\'{o}n, N. J.
Cerf, T.C. Ralph, J. H. Shapiro, S. Lloyd, Rev. Mod. Phys. \textbf{84}, 621 (2012).

\bibitem{Lev1}
A. Leverrier, Phys. Rev. Lett. \textbf{114}, 070501 (2015).

\bibitem{Lev2}
A. Leverrier, Phys. Rev. Lett. \textbf{118}, 200501 (2017).

\bibitem{FF1}
F. Furrer, T. Franz, M. Berta, A. Leverrier, V. B. Scholz, M. Tomamichel, R. F. Werner,
Phys. Rev. Lett. {\bf 109}, 100502 (2012), Phys. Rev. Lett. {\bf 112}, 019902(E) (2014).

\bibitem{FF2}
F. Furrer,
Phys. Rev. A {\bf 90}, 042325 (2014).

\bibitem{FF3}
M. Berta, F. Furrer, V. B. Scholz,
J. Math. Phys. {\bf 57}, 015213 (2016).

\bibitem{3MDI}
Y. Wu, J. Zhou, X. Gong, Y. Guo, Z.-M. Zhang, G. He, 
Phys. Rev. A {\bf 93}, 022325 (2016).

\bibitem{multik}
C. Ottaviani, C. Lupo, R. Laurenza, S. Pirandola,
arXiv: 1709.06988 (2017).

\bibitem{NOTAvacuum}
We put $\hbar = 1$ and assume the commutation relations of the form
$[ q , p ] = 2i$ \cite{paris}. In this way the output of homodyne
detection over the vacuum state is a Gaussian variable with variance $1$.

\bibitem{paris}
A. Ferraro, S. Olivares, and M. G. A. Paris,
{\it Gaussian States in Quantum Information}
(Bibliopolis, Napoli, 2005).

\bibitem{io}
C. Lupo, C. Ottaviani, P. Papanastasiou, S. Pirandola,
arXiv: 1712.00743 (2017).

\bibitem{NOTAmean}
For simplicity, here we have assumed that
$\langle q_Z \rangle = \langle p_Z \rangle = 0$.
Otherwise, one must replace
$\langle q_Z^2 \rangle \to \langle q_Z^2 \rangle - \langle q_Z \rangle^2$
and
$\langle p_Z^2 \rangle \to \langle p_Z^2 \rangle - \langle p_Z \rangle^2$.

\bibitem{NOTApergliidioti}
It is obvious that they still require to communicate the entries of the CM that have been
locally estimated. The CM, however, only contains a negligible amount of information about the
local raw keys.

\bibitem{Pacher}
N. Walenta, A. Burg, D. Caselunghe, J. Constantin, N. Gisin, O. Guinnard, R. Houlmann, P. Junod, B. Korzh, N. Kulesza,
New J. Phys. {\bf 16}, 013047 (2014).

\bibitem{Lev2010}
A. Leverrier, F. Grosshans, P. Grangier,
Phys. Rev. A {\bf 81}, 062343 (2010).

\bibitem{Jouguet2012}
P. Jouguet, S. Kunz-Jacques, E. Diamanti, A. Leverrier,
Phys. Rev. A 86, 032309 (2012).

\bibitem{Usenko}
L. Ruppert, V. C. Usenko, R. Filip,
Phys. Rev. A {\bf 90}, 062310 (2014).

\bibitem{Panos}
P. Papanastasiou, C. Ottaviani, S. Pirandola,
Phys. Rev. A {\bf 96}, 042332 (2017).

\bibitem{China1}
X. Zhang, Y. Zhang, Y. Zhao, X. Wang, S. Yu, H. Guo,
Phys. Rev. A {\bf 96}, 042334 (2017).

\bibitem{NOTAadc}
As a matter of fact, the application of the ADC 
introduce a quantization error and does not
preserve Gaussianity. 
The effects on the quantization error in parameter
estimation will be discussed in details in a future work.

\bibitem{Renner}
R. Renner,
Ph.D. thesis,
Swiss Federal Institute of Technology (ETH) Zurich, 2005,
arXiv:0512258 (2005).

\bibitem{Tomamichel}
M. Tomamichel
Ph.D. thesis,
Swiss Federal Institute of Technology (ETH) Zurich, 2012,
arXiv:1203.2142 (2012).

\bibitem{Wolf}
M. M. Wolf, G. Giedke, and J. I. Cirac,
Phys. Rev. Lett. {\bf 96}, 080502 (2006).

\bibitem{Raul}
R. Garc\,ia-Patr\'on and N. J. Cerf,
Phys. Rev. Lett. {\bf 97}, 190503 (2006).

\bibitem{NOTAtypo}
Notice that Theorem 1 of \cite{Lev1} contains a typographical error,
where $4 (d+1) \log_2{2/\epsilon^2}$ appears instead of
$4 (d+1) \sqrt{\log_2{2/\epsilon^2}}$ in the expression for $\Delta_\mathrm{AEP}^{(1)}$.
This typo is introduced in Eq.\ (56) of the Supplementary Information of \cite{Lev1}.

\bibitem{Beta1}
P. Jouguet, S. Kunz-Jacques, and A. Leverrier,
Phys. Rev. A {\bf 84}, 062317 (2011).

\bibitem{Beta2}
P. Jouguet, S. Kunz-Jacques, A. Leverrier, P. Grangier, E. Diamanti,
Nature Photon. {\bf 7}, 378 (2013).

\bibitem{Beta3}
M. Milicevic, C. Feng, L. M. Zhang, P. G. Gulak,
arXiv:1702.07740 (2017).

\bibitem{Beta4}
X. Wang, Y.-C. Zhang, Z. Li, B. Xu, S. Yu, H. Guo,
arXiv:1703.04916 (2017).

\bibitem{China2}
Y. Zhao, Y.-C. Zhang, B. Xu, S. Yu, H. Guo,
arXiv: 1711.04225 (2017).


\bibitem{TCC}
R. Renner and R. K\"onig,
Second Theory of Cryptography Conference TCC,
Lecture Notes in Computer Science Vol. 3378
(Springer, New York, 2005), pp. 407--425.

\bibitem{Composable}
R. K\"onig, R. Renner, A. Bariska, and U. Maurer,
Phys. Rev. Lett. {\bf 98}, 140502 (2007).

\bibitem{Helstrom}
C. W. Helstrom,
Quantum Detection and Estimation Theory (New York: Academic, 1976)

\bibitem{NOTA-a}
More recently, the purified distance $P(\rho,\rho_0)$ is employed
instead of the trace distance \cite{Tomamichel}.
However, since $D(\rho,\rho_0) \leq P(\rho,\rho_0)$ we still have
$p_e \geq \frac{1}{2}\left( 1 - D(\rho,\rho_0) \right)
\geq \frac{1}{2}\left( 1 - P(\rho,\rho_0) \right)$.

\end{thebibliography}
\end{document}